\documentclass{ws-ijmpe}
\usepackage[super,compress]{cite}
\usepackage{url}
\usepackage[all]{xy}
\usepackage{amsmath}
\usepackage{amssymb}
\usepackage{cancel}
\usepackage{mathrsfs}

\begin{document}

\markboth{Xiao-Yan Wang, Xiang-Jun Chen}{ Neutrino Oscillation Caused by Local Symmetry Broken}


%%%%%%%%%%%%%%%%%%%%%%%%%%%%%%%%%%%%%%%%%%%%%%%%%%%%%%%%%%%%%%%%%%%%

\title{Neutrino Oscillation Caused by Local Symmetry Broken}

\author{Xiao-Yan Wang\footnote{Corresponding author}}

\address{ Physics Department, Sichuan University of Arts and Science, 
Dazhou City, 635000, China\\
wangxy20220621@163.com}

\address{ Physics Department, Harbin Institute of Technology, 
Harbin City, 150001, China\\
wangxy20220621@163.com}

\author{Xiang-Jun Chen}

\address{Physics Department, Harbin Institute of Technology, 
Harbin City, 150001, China\\
chenxj@hit.edu.cn}

\maketitle

\begin{abstract}
Neutrino flavor conversion is assumed to be induced by right-handed neutrino flavor conversion via seesaw mechanism. Neutrino oscillation is the macroscopic phenomenon before all neutrino flavor-flip interactions reaching equilibrium. The model for the hypothesis is the extension by introducing large mass Majorana right-handed  neutrino into the Standard Model and exerting horizontal symmetry only on the right-handed neutrino sector.
\end{abstract}


\ccode{PACS numbers: 02.20.-a, 11.15.-q, 12.15.-y, 13.15.+g}

%\tableofcontents

\section{Introduction}

Neutrino oscillation is considered due to flavor mixing with three mass eigenstates. The detection experiments of dark matter\cite{1,2} indicate that there does not exist light dark matter,  which were thought to be neutrino mass eigenstates. In this paper, we present a direct scenario that neutrino flavor conversion is  induced by flavor-flip  interactions between  Majorana right-handed neutrinos due to horizontal symmetry via seesaw mechanism. In this scenario, neutrino oscillation is macroscopic phenomenon before all flavor-flip neutrino interactions reaching equilibrium, which will lead to apparent neutrino chaos of random matrix \cite{3,4}.  

To provide the direct scenario and give neutrino tiny mass, we  establish an extension which introduces large mass Majorana right-handed neutrino into the Standard Model and exerts horizontal symmetry only on right-handed neutrino sector. Experiments indicate that the intensity of neutrino oscillation is comparable to the strength of weak interaction. Thus we arrange horizontal symmetry on right-handed neutrino sector spontaneously broken at energy scale a little higher than that of the standard model (SM), which will provide related gauge boson $ F_{\mu}$  with mass $M_F \gtrsim M_Z $ as well as right-handed neutrino mass $\sim M_Z, M_W$. The right-handed neutrino mass is in the range indicated by WIMP model ($GeV \sim TeV$).

In order not to generate unallowable lepton-number-violating interactions and extra anomalies, we will arrange horizontal symmetry only in right-handed neutrino sector which only can generate flavor-flip right-handed neutrino interaction. Neutrino oscillation could be induced via seesaw  mechanism\cite{5,6,7,8,9,10} which allows neutrino constantly hopping between left-handed state and right-handed state during flight. Right-handed neutrinos are considered as components of WIMP  which fills up the space around us.

This direct scenario also indicates that neutrino mass and oscillation vanish in the limit of completely broken extra horizontal symmetry, which means no extra horizontal symmetry no neutrino mass and oscillation. When the symmetry is broken, neutrino mass and oscillation are related through seesaw mechanism
\begin{eqnarray}
m_v \propto 1/M_R \propto 1/M_F \propto 1 /\langle \Delta_R\rangle.
\end{eqnarray}
         
When  $ \langle \Delta_R\rangle \rightarrow \infty  $, symmetry   $SU(2)_N$ is completely broken and  $M_R,  M_F \rightarrow \infty  $, then neutrino mass vanishes and oscillation which is induced by $F_{\mu}$-intermediating interactions   disappears. 

    Moreover, since extra boson mass is given by extra symmetry spontaneously broken, according to G. 't Hooft theory the extra part is renormalizable. Thus, our extension is composed of two renormalizable parts: the standard model (SM) and the extra part, and we assume this extension is renormalizable.

    In the latter of the paper, we will predict neutrino flavor transition probability for the model by calculating scattering cross section, which implies that the size of neutrino mixing angle is influenced by neutrino mass difference.

\section{Model}

As mentioned earlier, we introduce right-handed neutrino into the standard model (SM) on which an extra horizontal symmetry $SU(2)_N $ is exerted. Extra symmetry $SU(2)_N $ is spontaneously broken at energy scale of $\langle \Delta_R\rangle $ slightly above the scale of the standard model (SM), giving masses to right-handed neutrino  and corresponding intermediator $F^i_{\mu}(i=1\sim 3)$.

\subsection{Leptons}

Leptons are arranged as follows, omitting color degree (quarks are arranged as in the standard model (SM)) 
\begin{eqnarray}
 \binom{\nu}{l}_L =\binom{\nu_e}{e^-}_L,\binom{\nu_{\mu}}{\mu^-}_L \qquad  (2,-1)  , \nonumber \\
\nonumber \\
 l_R =e^-_R, \mu^-_R \qquad  (1,-2) , \nonumber \\
  \nu_R =\left( \nu_e , \nu_{\mu}\right)_R   \qquad   (1,0,2) .
\end{eqnarray}\\
In Eq. (2), quantum numbers in parentheses represent $SU(2)_L $ isospins and U(1)-hypercharges (2I+1,Y). The third quantum number of right-handed neutrino represents  isospin of extra horizontal symmetry $SU(2)_N $,  $2I_N+1  $.

The charge relationship is still 
\begin{eqnarray}
\mathcal{Q} = T_{3L} + \frac{Y}{2},
\end{eqnarray}
because horizontal interaction does not transfer electrical charge. 

There are two scalar fields responsible for spontaneously breaking the entire model to $U(1)_{em}$, which are 
\begin{eqnarray}
\varphi = \binom{\phi^+}{\phi^0}& \langle \varphi\rangle 
 = \binom{0}{v} &  (2,  +1) , \nonumber \\
&& \nonumber \\
\Delta_R = \left(\begin{array}{cc} \Delta_{11} & \Delta_{12}  \\ \Delta_{21}   &    \Delta_{22} \end{array} \right) & \langle \Delta_R\rangle  = \left(\begin{array}{cc} 0 & c  \\ d   &    0 \end{array} \right)  & (1, 0,3) .
\end{eqnarray}
The third quantum number of $\Delta_R $ field in Eq. (4) is still $SU(2)_N $ isospin $2I_N+1  $. 

We assign extra symmetry $SU(2)_N $  spontaneously broken at slightly above energy scale of the standard model $\langle \Delta_R\rangle \gtrsim \langle \varphi\rangle $, giving right-handed neutrino large Majorana mass and related boson $F_{\mu}$ mass; While VEV $\langle \varphi\rangle $ breaks the standard model to $U(1)_{em} $, giving both charged leptons and neutrinos Dirac masses.

So, in our model, there are 7 bosons $W^{\pm }_L, Z^0,  F^i_{\mu} (i = 1 \sim 3) $ and one photon $A_{\mu} $. Extra  3 bosons $F^i_{\mu}(i =1\sim 3) $ are associated with extra symmetry $SU(2)_N $.

\subsection{Boson masses}

The covariant derivatives of scalar fields are 
\begin{eqnarray}
 D_{\mu}\Delta_{Ra } =  \partial_{\mu}\Delta_{Ra } + g_F \epsilon^{abc} F^b \Delta_{Rc } + \cdots ,  \nonumber  \\
 \nonumber \\
D_{\mu}\varphi = (\partial_{\mu} - i\frac{1}{2}g_Y Y_{\mu} -i\frac{1}{2}g_w \tau^i_w W^i_{\mu})\varphi+ \cdots .
\end{eqnarray}
Due to rotation invariance of vacuum, we rotate scalar field $ \Delta_R $ so that the VEV becomes  
\begin{eqnarray}
\langle \Delta_R\rangle &= &\Big(0 \quad 0 \quad \langle \Delta\rangle \Big) ,
\end{eqnarray}
where $ \langle \Delta\rangle\sim c \sim d \gtrsim \langle \varphi\rangle $. Thus after extra symmetry $SU(2)_N $ is broken by $\langle\Delta_R \rangle $, the standard model (SM) is broken by $\langle\varphi\rangle $, so that boson mass Lagrangian is 
\begin{eqnarray}
\mathscr{L}_m^B&=&\frac{v^2}{4}\big \{  g^2_w(W^1)^2+ g^2_w(W^2)^2+ (-g_wW^3+g_Y Y)^2  \big \} \nonumber \\
&& \nonumber \\
&&+g^2_F\langle\Delta\rangle^2 F_1^2+g^2_F\langle\Delta\rangle^2 F_2^2 .
\end{eqnarray}
Eq. (7) shows 5 massive bosons and 2 zero-mass bosons $ F_3, A_{\mu} $, where $A_{\mu} $ is a photon. We will not discuss here  extra zero-mass boson $ F_3$, and we will find that it is irrelevant to flavor-flip horizontal interaction which we want to study.

Redefine 
\begin{eqnarray}
F^{\pm}&=&\frac{1}{ \sqrt{2} }\left( F_1 \pm i F_2     \right) .
\end{eqnarray}
So masses of these two bosons are 
\begin{eqnarray}
M_{F^{\pm}}&=&\sqrt{2}g_F \langle\Delta\rangle ,
\end{eqnarray}                       
where $ \langle \Delta\rangle\sim c \sim d \gtrsim v \sim O(10^2 \textrm{GeV}) $. Therefore, we assume $ M_F \gtrsim M_Z $.

\subsection{Neutrino mass}

Neutrino mass is produced by following Yukawa coupling 
\begin{eqnarray}
\mathcal{L}& =& - h^f_{\nu} \bar \psi^f_L \tilde{\varphi}\nu_R^f  - \frac{1}{2}h_{ab}^R \overline{H^c_{Ra}  }Ci\tau_2 \Delta_R H_{Rb} + h.c.  ,  \\
\tilde{\varphi}&=&i\tau_2\varphi^*  , \nonumber  \\
&& \nonumber  \\
\psi^f_L&=&\binom{\nu_e}{e^-}_L,\binom{\nu_{\mu}}{\mu^-}_L ;  \
 \nu_R^f=\nu_{eR},\nu_{\mu R};\ H_{\nu R}=\left(\nu_e \quad \nu_{\mu} \right)^T_R  . \nonumber
\end{eqnarray}
where superscript $f$ represents family. The first term of Eq. (10) gives Dirac mass of neutrinos $m_D^f=h_{\nu}^f \nu  $; The second term gives right-handed neutrino Majorana masses $ M_{\nu_{eR}}=\frac{1}{2}h^R_{11}d, M_{\nu_{\mu R}}=\frac{1}{2}h^R_{22}c   $. Assume $m_D^f \sim m_l^f $ (subscript $l$ represents charged lepton), $M_{\nu_{eR}} \sim M_ Z  $ and $m_D^f \ll M_R^f $, then seesaw relation is 
\begin{eqnarray}
m_{\nu_e}\approx -m^2_{\nu_{eD}}\bigg/M_{\nu_{eR}}  
= - (h^e_{\nu})^2v^2 \Bigg/\frac{1}{2}h_{11}^Rd   
\propto 1/m_F .
\end{eqnarray}

Eq. (11) can be written as 
\begin{eqnarray}
m_{\nu_e} \approx  \frac{(h_{\nu}^e)^2v^2   }{ M_{\nu_{eR}}   }  
\sim  \frac{(h_{\nu}^e)^2v^2   }{ M_Z  } .
\end{eqnarray}
If we choose coupling constant $h_{\nu}^e \sim 2 \times 10^{-6}  $ (electron coupling constant $h^e \sim 10^{-6} $), we can obtain experimental mass $m_{\nu_e}\sim 1 \textrm{eV} $. Thus, our model also implies that neutrino Dirac mass is the same order of charged lepton mass.

\section{Horizontal interactions and neutrino oscillations}

The dynamics Lagrangian of leptons is 
\begin{eqnarray}
\mathscr{L}_l &=& i \overline{\psi_{Ra}  }\gamma^{\mu} (\partial_{\mu} + ig_Y Y_{\mu} )\psi_{Ra}
+i \overline{\psi_L^f  }\gamma^{\mu}(\partial_{\mu} + i\frac{1}{2}g_Y Y_{\mu} -i\frac{1}{2}g_w \tau^i_w W^i_{\mu})\psi_L^f   \nonumber \\
&&-i\bar H_{\nu R  }\gamma^{\mu}i\frac{1}{2}g_F \tau^i_F F^i_{\mu}H_{\nu R}   ,
\end{eqnarray}
where
\begin{eqnarray}
\psi_L=\binom{\nu_e}{e^-}_L,\binom{\nu_{\mu}}{\mu^-}_L  ;   
\ \psi_R =\nu_{eR},\nu_{\mu R}, e^-_R, \mu^-_R ; 
\ H_{\nu(R)}=&\binom{\nu_e}{\nu_{\mu}}_{R}  . \nonumber
\end{eqnarray}
And subscript $a$ represents all kinds of leptons $a=\nu_e,e^-,\nu_{\mu}, \mu^- $; $\nu$ represents for neutrino; and superscript $f$ represents family. According to Lagrangian (13), we can obtain following horizontal interactions between right-handed neutrinos
\begin{eqnarray}
\mathscr{L}_{int}=g_F\left( \frac{1}{\sqrt{2}}F^-_{\mu} \bar \nu_{eR} \gamma^{\mu} \nu_{\mu R} +  \frac{1}{\sqrt{2}}F^+_{\mu} \bar \nu_{\mu R} \gamma^{\mu} \nu_{e R}                 \right) .
\end{eqnarray}
In Eq. (14), we have used Eq. (8). Superscripts $+,-$ represent muon-number which boson $F_{\mu}$ carrying. Since right-handed neutrino $\nu_R $ is Majorana particle with 0-lepton-number, it is proper to assume that right-handed neutrino cannot discern muon-number namely flavor state, and thus undiscriminately couples to boson $F^{\pm}_{\mu}$, so that horizontal flavor-flip interactions between right-handed neutrinos intermediated by  boson $F_{\mu}$ can be written as 
\begin{eqnarray}
g_F F_{\mu}\left( \frac{1}{\sqrt{2}} \bar \nu_{eR} \gamma^{\mu} \nu_{\mu R} +  \frac{1}{\sqrt{2}} \bar \nu_{\mu R} \gamma^{\mu} \nu_{e R}                 \right) .
\end{eqnarray}
Eq. (15) indicates that right-handed neutrino cannot recognize and flavor-indistinguishably couples $F^{\pm }_{\mu}$. 
Neutrino mass state generated by seesaw mechanism is superposition of $\nu^c_L$ and $\nu_R $, so there exists spontaneous transition of $\nu^c_L \leftrightarrows \nu_R $, and transition probability of $\nu^c_L \rightarrow \nu_R $ is far lower than probability of $\nu_R \rightarrow \nu^c_L$. Neutrinos in this paper are Majorana particles, and we have $\nu^c_{L,R}\equiv \nu_{L,R} $. So interactions which cause neutrino oscillation can be written as 
\begin{eqnarray}
\xymatrix{ \nu_{Li}\rightarrow \nu_{Ri} \ar[r]    & \ar[r]  \ar@{~}[d]^{F_{\mu}}    &       \nu_{Rj}\rightarrow \nu_{Lj} \\    &&    } .
\end{eqnarray}
According to our model, Feynman diagram of neutrino flavor transformation in neutrino beam is
\begin{eqnarray}
\xymatrix{ \nu_{Li}\rightarrow \nu_{Ri} \ar[r]    &   \ar[r] \ar@{~}[d]^{F_{\mu}}    &        \nu_{Rj}\rightarrow \nu_{Lj} \\  \nu_{Li}\rightarrow \nu_{Ri} \ar[r]  &\ar[r]& \nu_{Rk}\rightarrow  \nu_{Lk}  .  } 
\end{eqnarray}

Seesaw mechanism means that neutrinos constantly jump to and for between left and right-hand states i.e. seesaw movement. For very long time observation of neutrino oscillation (several years), we assume that the small transition probability of $\nu_{Li} \rightarrow \nu_{Ri} $ would be dragged higher by the large transition probability of $\nu_{Rj} \rightarrow \nu_{Lj} $ and then these two probability would not influence the intensity of interaction (16), flavor conversion probability due to interaction (16) depends only on right-handed neutrinos horizontal interactions intermediated by boson $F_{\mu} $. In this way, diagram (17) can be simplified to diagram (18)
\begin{eqnarray}
\xymatrix{\nu_{Li} \ar[r]    &   \ar[r] \ar@{~}[d]^{F_{\mu}}    &       \nu_{Lj} \\   \nu_{Li} \ar[r]  &\ar[r]&  \nu_{Lk}  .  }
\end{eqnarray} 
Right-handed neutrino mass becomes neutrino's hidden mass, as seesaw mechanism implies. It is important to note here that diagram (18) is simplified form of neutrino-neutrino interaction over a very long observation time. The actual figure should be diagram (17).

\section{Neutrino flavor transition probability predicted by the model}
\subsection{ Neutrino flavor transition probability indicated by  the value of  scattering cross section}

If B particle beam strikes against stationary target composed of A particle, the definition of  cross section is 
\begin{eqnarray}
\sigma = \frac{N}{ n_B \cdot N_A } 
= \frac{1}{ n_B \cdot N_A } \int d^2b~ n_B P(b)  
= \frac{1}{ N_A } \int d^2b~  P(b) ,
\end{eqnarray}
where $N$ is total number of scattering cases during unit time, $N_A $ is total number of A particles, $n_B $ is flux density of B particles (particle number passing per unit area per unit time), $P(b) $ is probability function of scattering cases with respect to $b $. Assume that flux density $n_B $ is constant in the region of interactions.
The last equality in Eq. (19) shows that scattering cross section can be understood as  average scattering probability per unit time of each target particle. If beam cross section is 1$\textrm{cm}^2$, (since cross section formula (19) is based on the convention of the paraxial condition $b \sim 0 $, we choose unit area of $1\textrm{cm}^2 $ as beam cross section),  then we can find:
\begin{itemlist}
\item $n_B $---Total number of incident particle B per unit time
\item $N_A $--- Number of target particle A
\item $N$--- Number of scattering cases per unit time
\end{itemlist}
If we take a neutrino passing through one beam section as incident particle B, and other particles passing through the cross section at the same time as target particle A, thus  neutrino beam particle is both incident particle and target particle, so in unit time target particle number $N_A=n_B $. Thus, according to Eq. (19), the value of $\sigma $ is equal to average scattering probability per unit time of each neutrino, that is average flavor transition probability per unit time of each  neutrino or  average flavor transition probability of each  neutrino.

Thus we convert scattering cross section $\sigma $ to average flavor transition probability of a single  neutrino.

 It is important to note that the distance of scattering interaction  are the order of $\frac{1}{M_N}\sim \frac{1}{GeV}$.

\subsection{Neutrino flavor transition probability predicted by the model}

As described above, neutrino oscillation can be approximately assumed to be caused by diagram (18) after a long observation time. Since we only need to estimate neutrino oscillation probability according to the model, diagram (18) can be used, while   right-handed neutrino can be viewed as hidden neutrino mass. At the beginning of propagation, all neutrino-neutrino scattering can be approximately expressed as
\begin{equation}
\xymatrix{ \nu_{\alpha} \ar[r]^{p} & \ar[r]^k \ar@{~}[d]_{N} &  \nu_{\beta}\\
 \nu_{\alpha} \ar[r]_{p'} & \ar[r]_{k'} &  \nu_{\gamma} ,}   \qquad \alpha, \beta, \gamma = e, \mu, \tau  .
\end{equation}    
$ \nu_{\alpha} (\alpha, \beta,  \gamma = e ,\mu, \tau) $ is initial neutrino and $N $ is intermediate particle such as $W^{\pm   },Z^0, F_{\mu} $. Based on previous discussion, mass of $F_{\mu} $ is close to $M_Z$ in the standard model.

The scattering amplitude of diagram (20) is 
\begin{eqnarray}
iM = \frac{-ig^2_N  }{ M^2_N }  [ \bar u_{\nu_{\beta}}(k) \gamma^{\mu} (1-\gamma_5) u_{\nu_{\alpha}  }  ) (p) ]  
\times [ \bar u_{\nu_{\gamma}}(k') \gamma_{\mu} (1-\gamma_5) u_{\nu_{\alpha}  } (p') ]  .
\end{eqnarray}
For $m_{\nu_e  }\sim m_{\nu_{\mu}  }\sim m_{\nu_{\tau}  } \sim 0    $, in the center of mass reference frame, scattering cross section for diagram (20) is 
\begin{eqnarray}
\sigma_{CM} = \frac{ \overline{ |M|^2 }  }{ 16 \pi E^2_{CM }  } 
= \frac{4^3\times g^4_z  }{ 16\pi E^2_{CM } M^4_{N } } (k \cdot k' )(p \cdot p' )  
= \frac{4 \times g^4_z  }{ \pi E^2_{CM } M^4_{N } } (k \cdot k' )(p \cdot p' )  ,
\end{eqnarray}
where we assume that the size of $g_F$ is close to $g_Z$ in the standard model(SM). It should be noted here that Eq. (22) is established on the condition that $b\sim 0 $. The measurement of neutrino oscillation probability is generally obtained under paraxial condition, so Eq. (22) is always valid throughout the derivation of final neutrino flavor transition probability.

In the center of mass reference frame, we have 
\begin{eqnarray}
 k \cdot k'  = \frac{1}{2} (2 k\cdot k')    
= \frac{1}{2}[ (k+ k')^2 -(m^2_{\beta} + m^2_{\gamma})] 
= \frac{1}{2}[ E^2_{CM}-(m^2_{\beta} + m^2_{\gamma})]  ,
\end{eqnarray}
\begin{eqnarray}
 p \cdot p' =  \frac{1}{2} (2 p\cdot p')  
= \frac{1}{2} [(p+ p')^2 -2m^2_{\alpha}]  
= \frac{1}{2} [E^2_{CM}-2m^2_{\alpha}]  .
\end{eqnarray}
So we get scattering cross section in the center of mass reference frame as follows
\begin{eqnarray}
\sigma_{CM} &=&\frac{4\times g^4_z  }{ \pi E^2_{CM } M^4_{N } } (k \cdot k' )(p \cdot p' ) 
= \frac{ g^4_z \times [E^2_{CM}-(m^2_{\beta} + m^2_{\gamma})  ]\times [E^2_{CM}-2m^2_{\alpha}] }{ \pi E^2_{CM } M^4_{N } }  \nonumber \\
&\approx&  \frac{ g^4_z \times E^2_{CM} }{ \pi  M^4_{N } }  . 
\end{eqnarray}
Set 
\begin{eqnarray}
S =(p+ p')^2=(k+ k')^2 = E^2_{CM}   .   
\end{eqnarray}
Substitute Eq. (26) into Eq. (25) and we get cross section in terms of $s$
\begin{eqnarray}
\sigma \approx  \frac{g^4_z  }{ \pi  M^4_{z } }\times s  \qquad (M_N \sim M_Z  )  .
\end{eqnarray}
Take relative static reference frame of initial neutrino before collision, and we have 
\begin{eqnarray}
s = 2 p \cdot p'+ 2m^2_{\nu_{\alpha}}   
\approx 2 p \cdot p'    
 \approx 2 m_{ \nu_{\alpha}  } E_{ \nu_{\alpha}  }   .
\end{eqnarray}
So we get cross section as 
\begin{eqnarray}
\sigma  \approx  \frac{ g^4_z  }{ \pi  M^4_{z } }\times s  
\approx \frac{g^4_z \times 2 m_{ \nu_{\alpha}  } E_{ \nu_{\alpha}  }}{ \pi  M^4_{z } }  .
\end{eqnarray}
To estimate magnitude order of neutrino transition probability, we assume that neutrino energy $E_{\nu_e}=1 \textrm{GeV} $, and
\begin{eqnarray}
 m_{ \nu_e  } &\approx& 1 \textrm{eV} = 10^{-9} \textrm{GeV} ,  \nonumber \\
M_Z &\approx& 90 \textrm{GeV} ,  \nonumber \\
  g^2_Z &=& \frac{ M^2_{Z }     }{4 \sqrt{2}  } G_F,  \nonumber \\
 G_F &=& 10^{-5}/(\textrm{GeV})^2 .
\end{eqnarray}
Substitute above values into Eq. (29) and get 
\begin{eqnarray}
\sigma  \approx  \frac{10^{-19}  }{16 \pi  } (\textrm{GeV})^{-2}
= \frac{1 }{4\pi  }\times 10^{-47}  \textrm{cm}^{2} 
\sim  10^{-20}\textrm{mb}  .
\end{eqnarray}
Here we assume that $m_{\nu_{\alpha}} =m_{\nu_e}\sim 1 \textrm{eV}    $ and neutrino energy $E = 1 \textrm{GeV} $. In general, masses of the first two generations of neutrinos $\nu_e,\nu_{\mu}  $ are both considered much less than that of the third generation $\nu_{\tau} $, so when $\nu_{\tau} $ participates in the interaction, mass of $\nu_{\tau} $ in Eq. (25) cannot be ignored. Thus Eq. (29) is not suitable for these interactions, and their cross sections must be obtained directly from Eq. (25). What we want to know is whether flavor transition probability predicted by the model conforms to experiments, so we only need to estimate the transition probability of $\nu_e \rightarrow \nu_{\mu}$, ignoring details such as the effect of neutrino mass.

According to above discussion, the value of cross section $\sigma$ corresponds to average flavor transition probability of a single  neutrino, so the magnitude of neutrino flavor transition probability is 
\begin{eqnarray}
P(\nu_{\alpha} + \nu_{\alpha} \rightarrow \nu_{\beta} + \nu_{\gamma} ) \sim 10^{-47} ,
\end{eqnarray}
of which the interaction distance $\sim \frac{1}{M_N} \sim \frac{1}{GeV } $.

Thus, probability of neutrino flavor transition predicted by the model is the order as  
\begin{eqnarray}
P(\nu_{\alpha} \rightarrow \nu_{\beta} )\sim O( 10^{-47} )  .
\end{eqnarray}

\subsection{Flavor transition probability indicated by experiments}

Consider following formula 
\begin{eqnarray}
P(\nu_{ \alpha } \rightarrow \nu_{\beta  }  ) = \sin^2 (2 \theta_{ij} ) \sin^2 ( \frac{t}{L_0} )
=\sin^2 (2 \theta_{ij} ) \sin^2 ( \frac{x}{L_0} ) ,
\quad L_0 = \frac{ 4E }{ \Delta m^2_{ij} }
\end{eqnarray}   
In the  distance of neutrino flavor-flip interaction, Eq. (34) can be written as
\begin{eqnarray}
P(\nu_{ \alpha } \rightarrow \nu_{\beta  }  ) 
= \sin^2 (2 \theta_{ij} ) \sin^2 ( \frac{x}{L_0} ) 
 \approx \sin^2 (2 \theta_{ij} ) (\frac{x}{L_0})^2  ,
\end{eqnarray}   
where $ x \sim \frac{1}{M_N} \sim \frac{1}{1GeV } $, and we have 
\begin{eqnarray}
P(\nu_{ \alpha } \rightarrow \nu_{\beta  }  )= \sin^2 (2 \theta_{ij} )\Big(\frac{(1GeV^{-1})}{L_0 }\Big)^2 
= \sin^2 (2 \theta_{ij} )\Big(\frac{ \Delta m^2_{ij}\times (1GeV^{-1})}{ 4E }\Big)^2     .
\end{eqnarray}
To estimate the magnitude order of Eq. (36), we take a concrete example such as $\nu_e \rightarrow \nu_{\mu} $ and $E\sim 1GeV$. According to experimental data, we take 
\begin{eqnarray}
&&\Delta m^2_{12} \approx 8 \times 10^{-5} (eV)^2 =8 \times 10^{-23} (\textrm{GeV})^2 \nonumber\\
&&\sin^2(2\theta_{12} ) = 0.86   .    
\end{eqnarray}
Substitute Eq. (37) into Eq. (36) , we get
\begin{eqnarray}
P(\nu_{ e } \rightarrow \nu_{\mu  } ) \approx  \sin^2 (2 \theta_{12} )  ( \frac{8 \times 10^{-23}}{4} )^2  
\approx 3 \times 10^{-46}  .
\end{eqnarray}
Obviously, the value predicted by the model in Eq. (33) are very close to experimental data in Eq. (38).

\section{ Neutrino oscillation predicted by the model}
According to our model, the Lagrangian of neutrino oscillation can be approximately as ('approximately' is used because diagram (18) substitutes for diagram (17))
\begin{eqnarray}
\mathscr{L}_{int} = -ig_Z' \sum_{\alpha} \bar \nu_{\alpha}  \gamma^{\mu} (1-\gamma_5) \nu_{\alpha} Z_{\mu}    
-ig_F \sum_{\alpha \ne \beta} \bar \nu_{\beta}  \gamma^{\mu} (1-\gamma_5) \nu_{\alpha} F_{\mu}  ,
\end{eqnarray}
where $\alpha,\beta $ mean neutrino flavor $e,\mu, \tau $. As mentioned earlier, we have
\begin{eqnarray}
 g_Z& \sim &g_F   ,\\
M_Z &\sim& M_F    ;   \nonumber
\end{eqnarray}
so the total approximate valid Hamiltonian can be written as 
\begin{equation}
H_{int} = \frac{G_F'}{\sqrt{2}} \int dx^3 (\sum_i \bar{\nu'_i }_L \gamma_{\mu}  \nu_{iL})(\sum_j \bar{\nu'_j }_L \gamma^{\mu}  \nu_{jL})  .
\end{equation}
Here $\nu'$ is not pure state, but a mixture of flavor states, and $\nu_{iL},\nu_{jL} $ are initial flavor states. So neutrino mixing is just macro result of all interactions. During propagation of neutrino beam, neutrino flavor distribution will fluctuate  until all neutrino-neutrino interactions reach equilibrium. Neutrino oscillation is just the fluctuation of neutrino flavor distribution before reaching the equilibrium distribution.

\section{ Branch ratio  $\sigma(\mu^- \rightarrow e^-)/\sigma(\mu^- \rightarrow \nu_{\mu}) $ according to the model}
Calculate $\sigma(\mu^- \rightarrow e^-)/\sigma(\mu^- \rightarrow \nu_{\mu}) $ according to our model to see whether it is within allowed range. According to our model, the lowest order of reaction $\mu^- \rightarrow e^- $ is shown in Figure 1.

\begin{center}

\begin{figure}[th]

\centerline{\includegraphics[width=4.3cm]{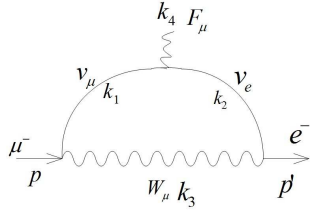}}
\caption{Lowest vertex of $\mu^- \rightarrow e^-     $ indicated by model}
\end{figure}

\end{center}

In Figure 1, for the sake of simplicity, we substitute approximate diagram (18) for diagram (17), and the scattering amplitude is 
\begin{eqnarray}
&&i\mathcal{M}(\mu^-\rightarrow e^-)   \nonumber \\
&&=\bar u_e(p')(-ig_w\gamma^{\alpha})\frac{-ig^{\alpha \beta}  }{M^2_W   }\frac{i \cancel{k}_2}{k^2_2   }
\times (-ig_F\gamma^{\eta})\frac{i \cancel{k}_1}{k^2_1   }(-ig_w\gamma^{\beta})u_{\mu}(p)F_{\eta}  .
\end{eqnarray}
Reaction $\mu^- \rightarrow \nu_{\mu} $ has the lowest order vertex 
\begin{eqnarray}
\xymatrix{ \mu^- \ar[r]^p
 &\ar[r]^{p'} \ar@{~}[d]^{W,k}&\nu_{\mu} .\\
& &  }
\end{eqnarray}              
The scattering amplitude is 
\begin{eqnarray}
i\mathcal{M}(\mu^-\rightarrow \nu_{\mu})=\bar u_{\nu_{\mu}}(p')(-ig_w\gamma^{\alpha})u_{\mu}(p)W_{\alpha}  .
\end{eqnarray}                     
In this way, we can get the branch ratio 
\begin{eqnarray}
\frac{\sigma(\mu^-\rightarrow e^-)  }{ \sigma(\mu^-\rightarrow \nu_{\mu}) }&\sim&
\frac{| \mathcal{M}(\mu^-\rightarrow e^-)  |^2    }{| \mathcal{M}(\mu^-\rightarrow \nu_{\mu}) |^2 }  
\sim  \frac{ g^4_w g^2_F }{ M^4_W  }\bigg/g^2_w 
=\frac{ g^2_w g^2_F }{ M^4_W  } \sim G^2_F/2  \nonumber \\
&<& 7\times 10^{-11}  , 
\end{eqnarray}
where, we assume $k_1\sim k_2 \sim O(\textrm{GeV}) $, and $g^2_W \sim g^2_F \sim M^2_W \frac{G_F}{\sqrt{2}} $.

The above branch ratio is just within allowable range \cite{11}. Because transition probability of $\nu_L \rightarrow \nu_R $ is so small, reaction $\mu^- \rightarrow e^ -$ is barely visible in the laboratory, and the branch will be smaller than (45).

\section{Conclusions}
We established an extension by introducing large mass Majorana right-handed neutrinos into the Standard Model and exerting horizontal symmetry only on the right-handed neutrino sector. We assigned the extra horizontal symmetry spontaneously broken at an energy level a little higher than the one of the Standard Model which provided right-handed neutrino Majorana mass comparable to the mass of Z or W and also provided related intermediators $F$ mass similar to the mass of Z or W. 

In the model, neutrino flavor conversion induced by the flavor-flip interactions between right-handed neutrinos via seesaw mechanism and neutrino oscillation was the macroscopic phenomenon before all flavor-flip interactions arrived at equilibrium which leading to apparent chaos of random matrix. 

The value of scattering section $\sigma$ could be expressed as average neutrino flavor  transition probablity of a single neutrino in beam. The neutrino flavor transition probability predicted by our model was consistent to the one indicated by experimental data.

\end{document}